\RequirePackage[2018-12-01]{latexrelease}
\documentclass{gGAF2e}

\begin{document}

\jvol{00} \jnum{00} \jyear{2024} 

\markboth{S. Pavlenko, E. Illarionov, D. Sokoloff}{Memory time fluctuations and instabilities in random media}


\title{MEMORY TIME FLUCTUATIONS AND INSTABILITIES IN RANDOM MEDIA}
\author{S. PAVLENKO{$^1$},
 E. ILLARIONOV$^{1,3}$ and D. SOKOLOFF$^{1,2,3\ast}$\thanks{$^\ast$Corresponding author. Email: sokoloff.dd@gmail.com}
\vspace{6pt}\\\vspace{6pt} 
{$^1$} Department  of Mechanics and Mathematics, Moscow State University, 119991, Moscow, Russia \\
{$^2$} Department of Physics, Moscow State University, 119991, Moscow, Russia \\
{$^3$} Moscow Center of Fundamental and Applied Mathematics, 119991, Moscow, Russia 
\\\vspace{6pt}{\today} }

\maketitle

\begin{abstract}
The role of memory time fluctuations for instabilities in random media is considered. It is shown that fluctuations can result in infinitely fast growth of statistical moments. The effect is demonstrated in the framework of light propagation in the Universe, which contains curvature fluctuations while remaining homogeneous and isotropic on average. 

\end{abstract}

\section{Introduction}

Instabilities in random media are important for many branches of science. To be specific, we consider this problem from the point of view of self-excitation of the magnetic field in turbulent and/or convective flows, i.e., in the framework of astrophysical dynamo. When considered in the framework of the Lagrangian approach (ideal MHD), the problem can be reduced to a set of ordinary differential equations

\begin{equation}
\frac{d {\bf B}}{dt} = {\bf B} \hat A,
\label{Lag}
\end{equation}
where $\bf B$ is the magnetic field vector considered as a row (rather than a column), the random matrix $\hat A$ consists of partial derivatives of the velocity field $v$, namely $\hat A = || \frac{\partial v_i}{\partial x_j}|||$, and we are interested in the magnetic field growth for $t \to \infty$.
Probability theory provides an efficient way to investigate this problem for various partial forms of the matrix random process $\hat A$ \citep[e.g.,][]{SI15, Ietal22}, known as Furstenberg theory \citep[a statement of this theory in a physics-friendly form can be found in][]{Zetal84}, and it looks reasonable to use this approach as far as possible. 
Of course, this is not the end of the story, and many important questions remain unclarified at this initial stage of research. It is sufficient to mention that this initial growth of the magnetic field can be combined with a catastrophic decrease of the spatial scale, which then leads to a abrupt decay of the magnetic field  \citep[e.g.,][]{Aetal81}. 
Another problem to be mentioned here is that the growth rate of the mean magnetic field (the first moment) is affected by turbulent
magnetic diffusion and microphysical magnetic diffusion due to electrical conductivity of plasma. The growth rate of the second moment of the magnetic field is also influenced by the microphysical magnetic diffusion, and the threshold of excitation of magnetic fluctuations depends on the magnetic diffusion.

Nevertheless, the problem seems interesting for understanding what happens during the development of the instability.

The key point in Furstenberg theory is that if a random matrix process loses memory at given times $t_n=\Delta_1 + \Delta_2 + \dots + \Delta_n$, where $n$ is the number of memory losses, then the problem can be reduced to the study of the product of many random matrices, which is quite accessible to study by methods of contemporary  probability theory. Of course, the initial idea here is to consider memory losses as equidistant and focus on the rate fluctuations between them. This is the approach used by various authors working in this area \citep[e.g.,][]{Eetal00, Ietal22}. 

Our point here is that the memory time in a real random flow can be viewed as a fluctuating quantity as well as the velocity between moments of memory loss. 
It seems realistic to consider the moments of memory loss $t_n$ as a Poisson flow of events, and the time intervals $\Delta_n$ as independent random variables distributed exponentially with mean $\Delta$. In the context of fluid mechanics, we can consider the vortex turnover time as a physical interpretation for the quantity $\Delta$.
There are several papers \citep[e.g.,][] {LS01, Eetal02, Eetal13} addressing this case, but they are written based on quite complicated forms of matrices $\hat A$, and new effects related to memory time fluctuations are not covered in them. 
The aim of this paper is to investigate the effects of memory time fluctuations using a few simple and instructive examples to highlight new effects arising in this problem.

\section{Product of random matrices}

Let the matrices of the random process $\hat A$ be time independent between memory loss moments $t_n$ and the random matrices $\hat A_n$ assigned to the time intervals between memory loss moments be statistically independent.
Then the solution of Eq.~\ref{Lag} at the time instant $t_n$ can be presented as a product

\begin{equation}{\bf B}_n = {\bf B}_0 \exp (\Delta_1 \hat A_1) \dots \exp (\Delta_n \hat A_n).
\label{prod}
\end{equation}
Here ${\bf B}_0$ is the initial condition while ${\bf B}_n$ is magnetic field at the time instant $t_n$ and the terms in the product of matrix exponentials are ordered according to the time intervals. 
The last point clarifies why it is pragmatic here to consider $\bf B$ as a row rather than a column. The matrices $\hat A_i$ denote realizations of the matrix random process $\hat A$ between the corresponding moments of memory loss.

Being interested in the growth of the mean magnetic field and taking into account that the matrices $\hat A_i$ with different indices $i$ are statistically independent, and assuming that $\hat A_i$ have the same statistical distributions, we immediately obtain

\begin{equation}\langle{\bf B}_n\rangle = \langle{\bf B}_0\rangle \langle\exp (\Delta \hat A)\rangle^n,
\label{first}
\end{equation}
where $\langle\dots\rangle$ stands for averaging. If we consider 
 only velocity fluctuations, and the time intervals $\Delta_i$ are the same, then ensemble averaging is performed only on the velocity.  If we consider memory loss moments as a Poisson flow of events, we have to include these fluctuations in the averaging. In any case, the problem of interest now reduces to averaging $\hat C = \langle\exp (\Delta \hat A)\rangle$ and computing the eigenvalues of the matrix $\hat C$. 

The crucial point here is that the matrices $\exp (\Delta_n \hat A_n)$ depend on $n$, and their mean $\hat
C=<\exp (\Delta \hat A)>$ is $n$-independent.

The investigation of the growth of the mean magnetic field is the simplest case of the Furstenberg theory. The calculation of the growth rates of higher statistical moments, including the second moment (dispersion), requires a more bulky algebraic construction \citep{IS21}, and to calculate the Lyapunov exponent responsible for the growth of typical realizations of the magnetic field, it is necessary to take into account that the product of matrices is noncommutative and to consider the integral equation for the invariant measure \citep{SI15}. 

The important point here is that the Lyapunov exponent turns out to be smaller than the growth rates of the statistical moments normalized by the corresponding moment number, and higher statistical moments grow faster than lower ones. This remarkable behavior, known as intermittency \citep{Zetal84}, occurs because individual, particularly fast-growing realizations can determine the growth of higher moments, even though their statistical weight decays exponentially fast.

At this point it becomes reasonable to separate the scalar factor $\det \hat A$, which in terms of dynamo theory means to consider incompressible flows, and to consider only matrices $\hat A$  with zero trace (${\rm Tr} \, \hat A=0$). Then the matrices $\hat C = \exp (\Delta \hat A)$ are unimodular, i.e., $\det \hat C =1$.

\section{A simple example: scalar field} \label{SE}

The above consideration is relevant for vector fields $\bf B$, but it can be slightly modified to include a scalar field as well. Then it becomes clear at what point the fluctuations of the memory time can be important. 

Consider a sequence of independent random numbers $a_n$ with a standard Gaussian distribution (mean is zero and unit variance).  Let $\Delta = {\rm const}$ and

\begin{equation}
c_n =e^{ a_1 \Delta} e^{ a_2 \Delta} \dots e^ {a_n \Delta}\,.
\label{exam}
\end{equation}
Straightforward calculations show that the corresponding Lyapunov exponent vanishes and $c_n$ grows subexponentially as $\exp {\sqrt n \zeta}$, where $\zeta$ is a standard Gaussian quantity. Even a very small diffusion being included in consideration destroy this subexponential growth. As for the mean value, it grows exponentially due to the intermittency effect. The growth rate of the first statistical moment is determined by $\langle e^{a \Delta}\rangle$, which is greater than 1. Note that the averaging takes place over the distribution of $a$. 

Now let $\Delta$ be an exponentially distributed random variable, i.e., the probability density of the probability that $\Delta = t_0$ scales as $\exp(-t_0/\tau)$, where $\tau$ is the mean value of $\Delta$. This means that the memory loss moments are taken from a Poisson flow of events. Now we need to average the distributions of $a$ and $\Delta$, and the latter diverges under the condition that $a$ can be larger (with non-decreasing probability) than $1/\tau$. Indeed, the contribution of such realizations to the mean scales as $\exp (a t_0 - t_0 /\tau)$ and the integral taken over $\tau_0$ diverges.

As for the rms value of $c_n$, the corresponding integral scales as $\exp (2 a t_0 - t_0/\tau)$ and diverges even more easily than when the mean value is determined. Of course, if the upper bound on the probability density function for $a$ is small enough to prevent divergence, the considered effect increases the growth rate in comparison equidistant memory loss times. 

We can say that very large time intervals without memory loss, which occur in a Poisson flow of events with exponentially small probability, do determine the mean values of $c_n$, since their contribution to the mean value grows faster than their statistical weight decreases.  Below we will show that this divergence is not a specific property of this illustrative example, but occurs for vector fields as well.    

\section{Vector field: Jacobi equation}

Let us now consider the effect described above for the vector field. We use here a useful model example proposed by \cite{Z64} in the context of the cosmological problem. The point is that an observer trying to measure the curvature of the spatial section of a homogeneous and isotropic on average Universe with vanishingly small mean curvature have to conclude that the Universe is open and its spatial curvature is negative.
This is because the separation of lines of light, i.e., geodesic lines, grows exponentially (rather than linearly, as it does in purely flat space), due to curvature fluctuations associated with particular celestial bodies \citep[see, e.g.,][for more details]{Letal05}.  The separation of geodesic lines is measured by the so-called Jacobi field or geodesic deviation $y$, which is governed by the Jacobi equation

\begin{equation}
y''+Ky =0.
\label{Jac}
\end{equation}
Here $K$ is the curvature (more precisely, the sectional curvature) computed at a given point on the main geodesic line, and the derivative is taken by the distance along this geodesic line. This distance can be considered as a spatial distance or, if desired, as time $t$.
If $K=0$, $y$ grows linearly, while for negative $K$ the geodesic deviation $y$ grows exponentially and oscillates for positive $K$. This test, originating from the early ideas of Gauss and Lobachevsky, provides a way for observational determination of the curvature of cosmological models. \cite{Z64} obtained that the geodesic deviation $y$ grows exponentially provided that $K$ is a random process with zero mean and nonvanishing fluctuations. 
In the context of this cosmological problem, we can consider $\Delta$ as the spatial scale at which the Universe is
homogeneous. Then $\Delta \approx 100$ megaparsec \citep{1983rav..book.....Z}.

This effect is highly appreciated in cosmology, but here it is interesting because the different growth rates of $y$ can be computed explicitly from a rather simple approach \citep{Setal21}.
For this purpose, let us rewrite Eq.~(\ref{Jac}) as a system of ordinary differential equations as 

\begin{equation}
{\bf z}' ={\bf  z} \hat A,
\label{Jacsys}
\end{equation}
where ${\bf z} = (y, \Delta y')$ and $ \hat A = \begin{pmatrix}
0 & - K \Delta \cr 1/\Delta & 0 
\end{pmatrix}$. Solving this system following Eqs.~(\ref{prod}, \ref{first}),
we obtain that the calculation of the growth rate for the first moment can be reduced to averaging of the random matrix $\begin{pmatrix} 
\cos \sqrt k & - \sqrt k \sin \sqrt k \cr \sin \sqrt k/\sqrt k & \cos \sqrt k
 \end{pmatrix}$ for $k=K\Delta^2$ and positive $K$,  and $\begin{pmatrix}
\cosh \sqrt k & \sqrt k \sinh \sqrt k \cr \sinh \sqrt k/\sqrt k & \cosh \sqrt k \end{pmatrix}$ for negative $K$. The averaging should be performed over the statistical distributions $K$ and $\Delta$. 
It can be seen that the same divergences appear here as in the previous section.

Our aim now is to compare two cases  in the framework of Eq.~(\ref{Jacsys}): $\Delta$ is a nonrandom constant
(say, $\Delta_0$) and $K$ fluctuates and $\Delta$ is an exponentially distributed random quantity with $\Delta_0$ as
the mean and $K$ fluctuates as in the previous case.

Calculating the growth rate of the second moment, we have to introduce the quantity $z_1z_1, z_1 z_2, z_2
z_1, z_2 z_2$, consider it as a vector $\bf q$ in the auxiliary 4D space, obtain the governing equation for $\bf q$
and perform the same calculations as above using matrices $4 \times 4$ instead of the matrices $2
\times 2$ \citep{IS23}.

\section{Results}

Following the method discussed above, we computed the growth rates of the first and second statistical moments for different probability distributions of the random curvature $K$. As expected, the result diverges if $K$ is less than a threshold $K^*_n$ (here $n=1$ for the first moment and $n=2$ for the second). The threshold for the first moment is 

\begin{equation}
K^*_1 = -1/\Delta^2
\label{thr1}
\end{equation}
and the threshold for the second moment is

\begin{equation}
K_2^* = -1/(4 \Delta^2).
\label{thr2}
\end{equation}
This result is exactly as expected from the simple example in section~\ref{SE}. 

Further investigation of the discussed effect is relevant for negative $K$, which is bounded from below by a value slightly larger than the threshold value. Therefore, we restrict ourselves to $K$ uniformly distributed in the interval $-\alpha K^*_n, 0$, where $n$ is chosen to correspond to the number of the moment under discussion and $\alpha$ is a constant close to 1.  Averaging the matrices and further calculation of growth rates requires simple numerical calculations, which do not need special discussion here.

\begin{figure}[!ht]
\begin{center}    \includegraphics[width=0.45\linewidth]{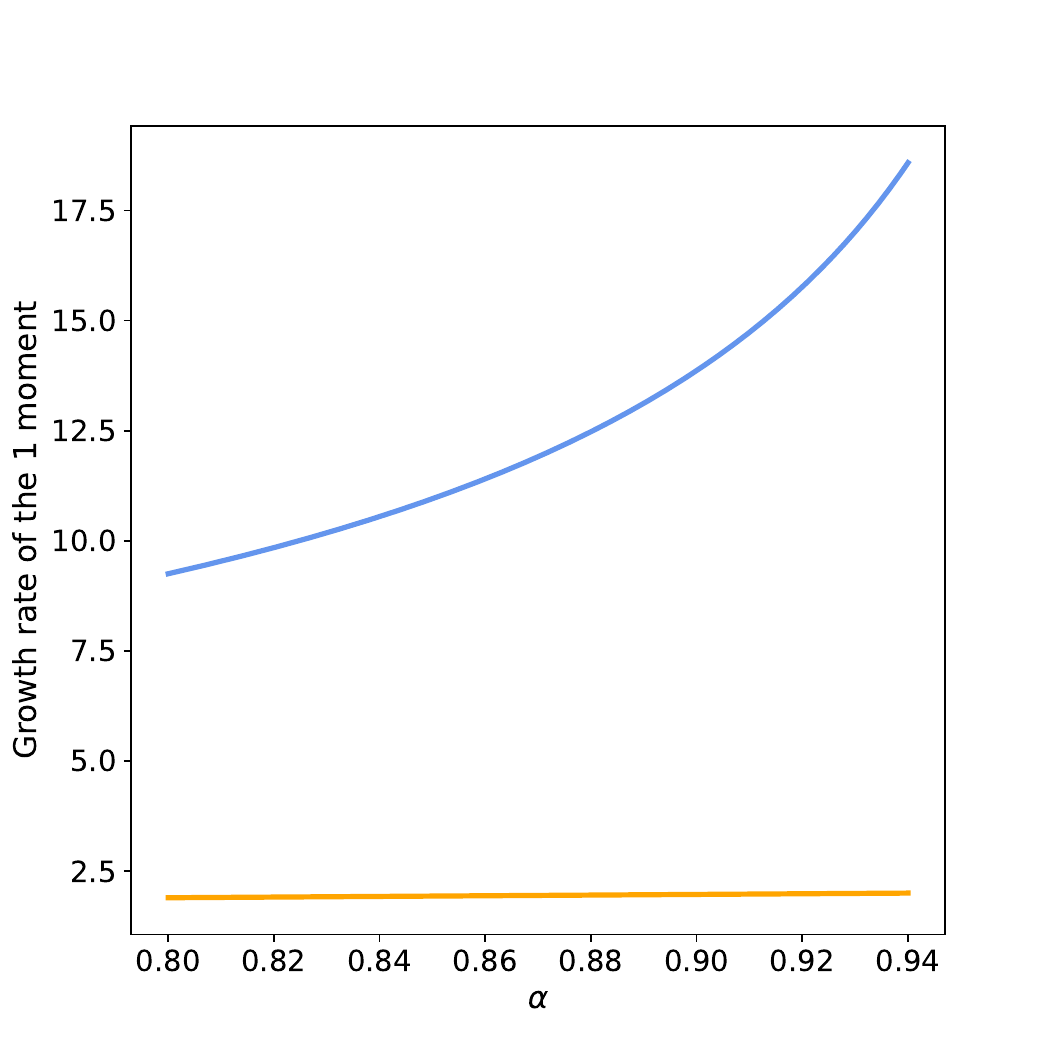}
 \includegraphics[width=0.45\linewidth]{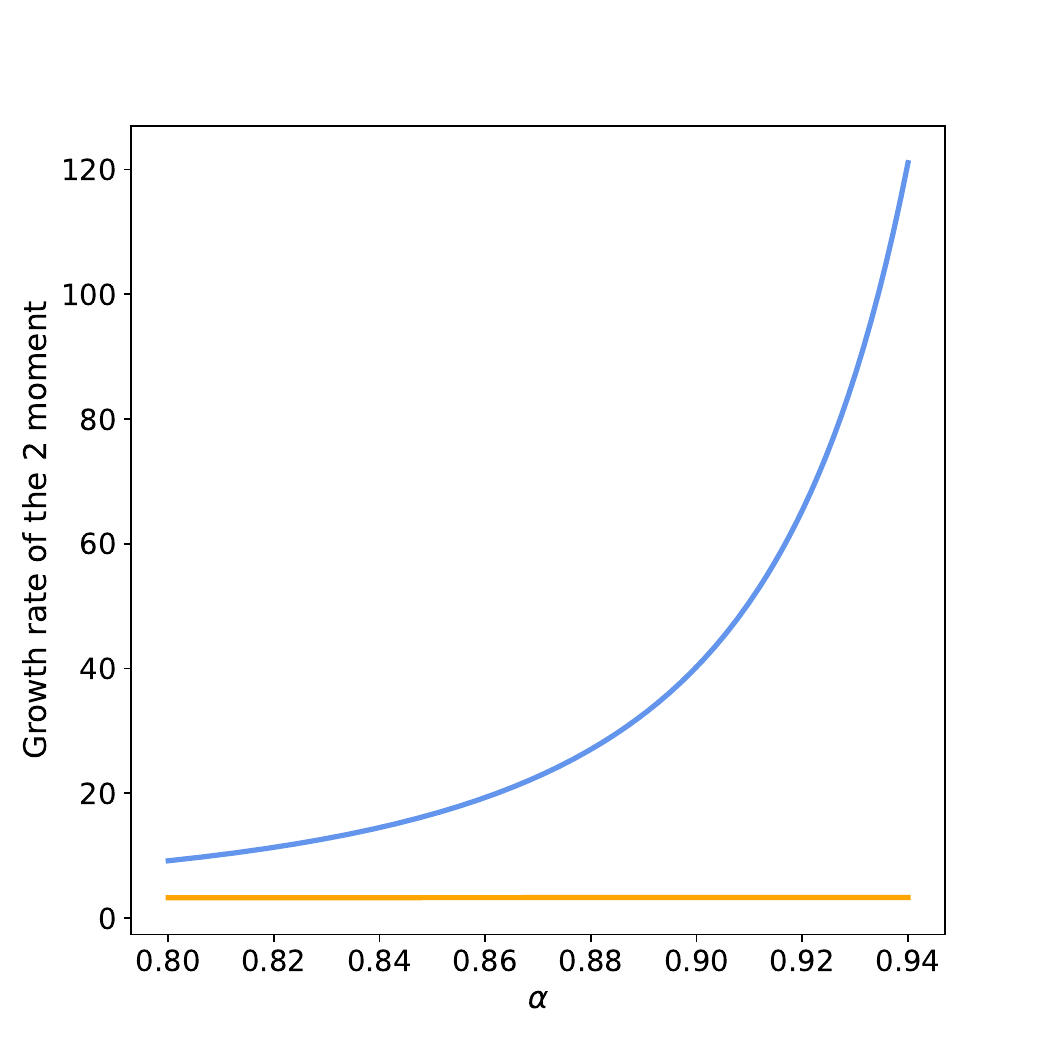}
 \end{center}
    \caption{Growth rate of the first moment (left panel) and the second moment (right panel) as a function of $\alpha$ for  the exponentially distributed $\Delta$ with 1 as its mean value (blue line) and $\Delta=1$ considered as a nonrandom quantity  (orange line).  }
	\label{F1}
\end{figure}

The growth rate of the first moment as a function of $\alpha$ is presented in the left panel of Fig.~\ref{F1} in
comparison with the result obtained for constant memory time. It can be seen that the memory time fluctuations contribute significantly to the growth rate of the first moment provided that the governing parameters are close to the threshold value.

The results for the second moment are shown in the right panel of Fig.~\ref{F1}. Comparing the results in both figures, we
conclude that the growth rates do indeed diverge, as expected when $\alpha \to 1-0$. The growth rate for
the second moment is larger than for the first moment. This is consistent with naive expectations.

\section{Conclusion and discussion}

Summarizing, we conclude that memory time fluctuations lead to a specific effect that occurs when $K$ becomes negative with a non-vanishing probability. From the equations~(\ref{thr1}, \ref{thr2}), thresholds arise, and the corresponding growth rates diverge close to the thresholds. We restrict ourselves to computing only growth rates, although it looks quite convincing that the statistical moments themselves also diverge in these situations. Again, we consider the first two statistical moments, but it is natural to expect that the higher moments diverge even more faster than the first ones.

In principle, the divergence of statistical moments can occur in various probability problems. A classical example here is the ratio of independent Gaussian random variables (the Cauchy distribution). Indeed, the denominator may with very small probability become a vanishing quantity, leading to a divergence of the mean. The effect associated with the Cauchy distribution poses certain problems in practical statistics. However, we can overcome it with the assumption that the very large deviations in the Cauchy distribution are unrealistic in practice. Perhaps the same can be said for the effect discussed here. The Poisson flow of events looks like a standard model for noise in memory time, but extremely large intervals between memory losses are still unrealistic in practice. In this sense, we can say that the effect highlighted in the paper should be considered as a limitation of the standard probabilistic models used by physics. 

The effect discussed can be compared to another divergence that occurs for instability in random media, e.g., \cite{Ketal23} demonstrate that explosive growth of the averaged variables happens if higher statistical moments of instability drivers diverge. However, the effect presented above appears much more dramatic. Indeed, the Poisson flow of events is a standard mathematical model for radioactive decay and is used for, say, radiocarbon dating. Statistical noise arising from the Poisson flow of events does propagate in radiocarbon data, but no one expects anything dramatic here. If the instability is taken into account, however, we do encounter a dramatic divergence.

We demonstrated and investigated the effect on a simple illustrative model, however, we have no reason to expect that the effect arises only for this model, at least the simulations in \cite{Oetal24} confirm this expectation.

\section*{Acknowledgements}

EI and DS acknowledge the the financial support of the Ministry of Education and Science of the Russian Federation as part of the program of the Moscow Center for Fundamental and Applied Mathematics under the agreement 075-15-2022-284.

\bibliographystyle{gGAF}
\bibliography{literature}
\vspace{12pt}

\end{document}